\documentclass[12pt]{article}
\usepackage{graphicx}

\date{}
\title{\bf 
Spin Dynamics Of $qqq$ Wave Function  On  Light Front In  High Momentum Limit Of QCD : 
 Role Of  $qqq$  Force }
\author{
{\normalsize\bf
A.N.Mitra \thanks{Email: (1)ganmitra@nde.vsnl.net.in;
(2)anmitra@physics.du.ac.in}
}\\
\normalsize 244 Tagore Park, Delhi-110009, India}
       
\begin{document}

\maketitle

\begin{abstract}
The contribution  of a spin-rich $qqq$  force ( in conjunction  with  pairwise $qq$ forces) to the analytical structure of the $qqq$ wave function   
is worked out  in the high momentum regime of QCD  where the confining interaction may be ignored, so that         
the dominant effect is $Coulombic$. A distinctive  feature of this study is that  the spin-rich  $qqq$  force is generated by  a $ggg$ vertex 
( a genuine part of the  QCD Lagrangian )  wherein  the 3 radiating gluon lines end on  as many quark lines, giving rise to a (Mercedes-Benz type)  
$Y$-shaped diagram.   The dynamics  is that of a  Salpeter-like equation  ( 3D support for the kernel)  formulated covariantly  on the light front, a la   Markov-Yukawa Transversality Principle (MYTP)  which warrants a 2-way interconnection between  the 3D and 4D Bethe-Salpeter (BSE)  forms   
for 2 as well as 3  fermion quarks. With these  ingredients,  the differential equation for the 3D wave function $\phi$  receives well-defined  
contributions from  the $qq$ and $qqq$ forces. In particular  a $negative$ eigenvalue of the spin operator  $i \sigma_1.\sigma_2\times \sigma_3$  
which is an  integral  part   of   the $qqq$ force,  causes a characteristic  singularity in the differential equation,  signalling the dynamical effect  
of a  spin-rich $qqq$ force  not yet  considered in the literature.  The potentially crucial role of this interesting effect   vis-a-vis  the so-called  
`spin anomaly' of the  proton, is  a subject of considerable physical  interest.     
\end{abstract}

\section{ Introduction}

The concept  of a fundamental  3-body force (on par with a 2-body force)  is  hard to realize in physics, leaving aside certain ad hoc representations 
of higher order effects, for example  those of $\Delta, N^*$ resonances  in hadron physics. At the deeper quark-gluon level on the other hand, a  
truly 3-body $qqq$ force  shows up as a folding of a  $ggg$ vertex ( a genuine part of the gluon Lagrangian in QCD) with 3 distinct ${\bar q}gq$ 
vertices, so as to form a  $Y$-shaped diagram (see fig 1 below).  Indeed  a 3-body $qqq$ force of this type, albeit  for `scalar' gluons,  was 
first suggested  by Ernest Ma \cite{ ErMa75},  when QCD was still  in its infancy. [ A similar representation is also possible for $NNN$ interaction via 
$\rho\rho\rho$  or $\sigma\sigma\sigma$ vertices, but was never in fashion in the literature  \cite{McK94}].  We note in passing that  a 
$Y$-shaped (Mercedes-Benz type) picture \cite{Mitr83}  was once  considered in the context of a preon model for quarks and leptons. 

\setcounter{equation}{0}
\renewcommand{\theequation}{1.\arabic{equation}}

In the context of QCD  as a Yang-Mills field,  a $ggg$ vertex has a momentum representation of the form \cite{Tay78}
\begin{equation}\label{1.1}
W_{ggg} =  -i g_s f_{abc} [(k_1-k_2)_\lambda \delta_{\mu\nu} + (k_2-k_3)_\mu \delta_{\nu\lambda} + (k_3-k_1)_\nu \delta_{\lambda\mu}]
\end{equation} 
where the 4-momenta emanating from the $ggg$ vertex satisfy $k_1 + k_2 + k_3 = 0$, and $f_{abc}$ is the color factor.  When  this vertex is folded  
into 3  ${\bar q}gq$ vertices of the respective forms $ g_s {\bar u}(p_1') i\gamma_\mu \{\lambda_1^a /2 \}  u(p_1)$, and two similar terms,  the 
resultant $qqq$ interaction matrix (suppressing the Dirac spinors for the 3 quarks)  becomes  
\begin{equation}\label{1.2} 
V_{qqq} = \frac{ g_s^4}{2^3} [ i\gamma^{(1)}. (k_2 - k_3)  \gamma^{(2)}. {\gamma^{(3)}  + \{2\} + \{3\} ]  \{\lambda_1\lambda_2\lambda_3\}  /\{ k_1^2 k_2^2 k_3^2} \} 
\end{equation} 
where $k_i = p_i -p_i' $;  $\lambda_i$   are the color  matrices  which get contracted into the 
corresponding scalar triple products  in an obvious notation. [Note that   the  flavour indices are absent  here since the quark gluon interaction is 
flavour blind].   
\par    
This interaction will  be considered in conjunction  with 3  pairs of $qq$ forces  within the framework of a Bethe-Salpeter type dynamics to be specified below.  Before proceeding  in this direction,  it is  in order  to explain a possible  motivation behind the use of  a direct $qqq$ force with such  a rich spin dependence.  Apart from the intrinsic beauty of this  term,  the immediate  provocation for its use comes  from  the  issue of "proton spin"   which,  after making  headlines about two decades ago, has  come to the fore once again, thanks to the progress of  experimental  techniques in polarized deep inelastic scattering off polarized protons, and their variations thereof,  which allow for an experimental  determination of  certain key   
QCD  parameters by relating them  to certain  observable quantities emanating from  external probes ; ( see a recent review  \cite{Bass05} for references and other details).   On the other hand it is also of considerable theoretical interest  to determine these very quantities directly  from the $intrinsic$ premises of QCD  $provided$  one has a " good " $qqq$ wave function  to play with.  Such a plea would have sounded rather utopian in the early days of QCD when phenomenology was the order of the day. Today however many aspects of QCD are understood well enough to make such studies worthwhile by hindsight, with possible ramifications beyond their  educational value.  The role of the direct 3-body force may be seen in this light,  while working for simplicity in the experimentally accessible  regime of valence quarks.  To that end  this paper  is specifically concerned with the  effect of the 3-body force (1.2) on the analytical structure of the $qqq$ wave function, while the formalism dealing with the contributions of the various operators ( $i\gamma_\mu \gamma_5$, 2-gluon effects, etc)  to the proton spin is  reserved  for a subsequent paper.    

\subsection{Theoretical Ingredients}

In the valence quark regime, we need to consider a $qqq$ system governed by pairwise $qq$ forces  as well as a direct 3-quark force of the type (1.2).  
A further simplification occurs in  the high momentum regime where the effect of confining forces may be neglected, so that only coulombic forces need       
be kept track of. We shall take  the dynamics of a $qqq$ hadron  in the $high- momentum$  regime to be  governed by   a Bethe-Salpeter Equation (BSE)  whose kernel is a sum  of three pairs of (coulombic) $qq$ forces plus a single 3-body term (1.2).    Unfortunately   the 4D form of     
the BSE  is too general to be of  practical value for $qqq$  dynamics, so a  better option is the Salpeter Equation \cite{Salp52} representing  its     
instantaneous form.  And,  except for its lack of covariance, the Salpeter  equation  has the remarkable property of  3D-4D interlinkage,     
a feature that had been present all along in the original formulation itself \cite{Salp52}, but had somehow remained  hidden from view in the literature, 
until   clarified   \cite{MiSo01} in  the context of a comprehensive $two-tier$  BSE formalism developed independently in a covariant manner \cite{CGSM89}. (A practical significance of the  `two-tier' formalism is that the  reduced 3D form  is ideal for the determination of  hadron mass spectra  of both     
$q{\bar q}$ and $qqq$  types \cite{SiMi88, Shar94},  while the  reconstructed 4D form   is convenient  for the evaluation of transition amplitudes     
\cite{SiMi88} via Feynman techniques for  loop diagrams; see also Munz et al ref.(9) ). A covariant formulation of the BSE  is centered around the hadron  4-momentum $P_\mu$  in accordance with  the Markov-Yukawa Transversality Principle (MYTP) \cite{Mark40, Yuka50}, which was shown     
to be  a `gauge principle'  in disguise \cite{LuOz77}. It ensures that   the interactions among the constituents  be  $transverse$ to the direction of     
$P_\mu$.   In the high momentum regime to be considered here,  the confining interaction has been ignored for simplicity,   which leaves the     
3D form of the BS dynamics  inadequate for   mass spectral determination, yet  its dynamical  on the spin-structure of the wave function   should  
be realistic enough for dealing  with  the  hadron spin  in the high momentum limit.    
\par
Now to another vital element  of the theory:  Although  the Salpeter Equation, as the instant form of BSE,  admits of a covariant formulation in the     
rest frame of the total hadron 4-momentum $P_\mu$,  it  suffers from  certain  ill-defined  
4D loop integrals due to a `Lorentz-mismatch' among the rest-frames of the 
participating hadronic composites, resulting in time-like momentum components 
in the exponential/gaussian factors associated with their vertex functions. 
This is especially true of  triangle loops, such as applicable to the pion form 
factor and $\rho-\pi\pi$ coupling \cite{Sant93} where this disease 
causes unwarranted "complexities" in the amplitudes, while  two-quark 
loops just escape this pathology. For a possible remedy  against   this 
disease, without losing  the benefit  of  an MYTP - based  3D-4D interconnection, a promising candidate is the {\it light-front}
 approach of Dirac \cite{Dira49} by virtue of its bigger $(7)$ stability group  compared with $6$ for the {\it {instant form}} 
theory. Its basic simplicity was first noticed by  Weinberg \cite{Wein66} who formulated the infinite momentum frame 
towards the same end. A covariant generalization  of MYTP on the light front,  requiring  the use of two null  4-vectors $n_\mu $ and ${\tilde n}_\mu$ 
that satisfy $n^2 = {\tilde n}^2 = 0$ and $ n . {\tilde n} =1$, cures the `Lorentz mismatch' disease noted above \cite{Mitr99}. The remaining problem of  
$n$- dependence of transition  amplitudes is solved through  a  simple ansatz of 
`Lorentz completion' \cite{Mitr99, Mitr00}, so as to  yield  a Lorentz-invariant  pion e.m. form factor in accord with experiment .  A similar 
approach had been considered by Carbonell et al \cite{Carb98} in the context of the Kadychevsky-Karmanov formalism \cite{ Kady68, Karm80},     
except for  their missing out on  the  second (dual) null-vector ${\tilde n}$ which  happens to be crucial for  recovering a Lorentz - invariant structure 
in a more natural way.

\subsection{ Plan of the Paper}  

The plan of the paper is  based on  an interlinked  3D-4D BSE  formalism characterized by  a Lorentz-covariant  3D support for its kernel a la MYTP  \cite{Mark40, Yuka50},  adapted to  the light front (LF) \cite{Mitr99}. The  full structure of  the kernel is a sum of 3 pairs of coulombic $qq$ forces plus a $qqq$ force, Eq.(1.2), whose 3D support is implied by the fact  that all internal momenta $q$  be $transverse$  to $P_\mu$ , viz.,   ${\hat q}_\mu = q_\mu-q.PP_\mu/P^2$. However  the propagators are left untouched in their standard 4D forms. [The  light front formulation requires  a $collinear$ frame \cite{Mitr99} which is further  specified in Section 2 ]. The strategy now lies in a step-wise  reduction of the (Salpeter-like) 4D  Master Equation involving  the actual  (4D)  fermionic wave function $\Psi$. Step A  consists in  expressing  $\Psi$  in terms of  an auxiliary  (bosonic)  4D quantity $\Phi$ satisfying an equivalent (bosonic) 4D (Salpeter-like) equation. Step B  involves a " Gordon reduction "  to eliminate the $\gamma_\mu$   matrices in favour of $\sigma_{\mu\nu}$ matrices. Finally Step C consists in  a  3D reduction of this `bosonized'  Master Equation, and a subsequent $reconstruction$ of the  original  
$\Psi$ in 4D form by a suitable reversal of steps.  In the process a  3D scalar wave function $\phi$ is introduced  which  not only facilitates an explicit solution of the 3D (albeit fully covariant) Salpeter-like equation  but is also a key component of the (reconstructed) 4D fermionic wave function $\Psi$.    
To  facilitate the process of 3D-4D interlinkage,  a Green's function approach is employed  a la \cite {Mita99},  from which it is straightforward to 
derive the corresponding wave functions via the appropriate `pole' limits.   The entire exercise involves  a close correspondence between the (earlier) instantaneous form \cite{Mita99} and the (later)  LF form \cite{Mitr99}, so as  to  project only the latter  by making  free use  of the results of the  former. 
A new element  will be a generalization of  the  earlier formalism \cite{Mita99} so as to include the effect of the 3-body force, Eq (1.2),  on the structure  of the $qqq$ Green's function which will require a more elaborate strategy, keeping in view the relative strengths of $qq$ and $qqq$ forces.         
\par
After a short account of the correspondence between the instant and LF forms of the dynamics, {\bf Section 2} is devoted to  Steps A and B for converting  the Master Equation for the actual 4D $qqq$ wave function $\Psi$ (fermionic) to  an equivalent  4D (bosonic) form involving  $\Phi$,  with all $\gamma$-matrices eliminated in favour of Pauli matrices, by exploiting  the effective 3D support for the  $qq$ and $qqq$ kernels on the light-front, as described above. {\bf Section 3} describes the first part of Step C, viz., the use of   $Green's$- function method for the 3D reduction  of the Master Equation for the 4D Green's function $G$ corresponding to $\Phi$, resulting in an integral equation for the 3D Green's function ${\hat G}$ corresponding to $phi$, specialized to the large momentum regime on the light-front.  The second part of Step C, namely its  reconstruction  back to  the  4D  quantity $G$  a la \cite{Mita99},  leading  to a formal  connection between $\Phi$ and $\phi$, is the subject of {\bf Section 4}, now with the added complexity behind the  inclusion of the 3-body $qqq$ force along with the (dominant) pairwise $qq$ forces. [ The subsequent reconstruction of  the 4D  fermionic form $\Psi$, follows from the results of Section 2]. As a further aid to the understanding of the 3D-4D interconnection, Appendix A describes the complete procedure for a prototype 
$qq$  subsystem, whose results are freely used for justifying several steps in Sections 3 and 4. In {\bf Section 5},  the  full
3D BSE for $\phi$ is set up in a simplified form designed to  check for the  dynamical effect of spin present in both the $qq$ and $qqq$ forces.  on the structure of  the differential equation for $\phi$  (the dominant effect being of the latter ! ).   {\bf Section 6} is devoted to a short critique of the role of the  $V_{qqq}$ term on the analytic  srtucture of the $qqq$ wave function, while the derivation of the requisite  QCD parameters for  the baryon from the $forward$ scattering amplitude  off the $i\gamma_\mu \gamma_5$  operator, via  the quark constituents, as well as  a more elaborate  2-gluon  contribution to the proton spin,  is reserved for a subsequent paper. 

\section{ Salpeter Eq on LF : From  $\Psi$ To  $\Phi$} 

\setcounter{equation}{0}
\renewcommand{\theequation}{2.\arabic{equation}}

\subsection{ Master Eq : Instant Vs LF Forms of Dynamics}

Since the first step in the formulation of  a covariant Salpeter-like equation on the light front (LF) is to establish a correspondence between the instant and LF forms of the dynamics, we first  recall  some definitions \cite{Mitr99} for the LF quantities  $p_{\pm} = p_0 \pm p_3$ defined covariantly as  $p_+ = n.p \sqrt{2}$ and $p_- = -{\tilde n}.p\sqrt{2}$. while the perpendicular components     
continue to be denoted by $p_\perp$ in both notations.  Now  for  a typical  internal momentum $q_\mu$,  the  parallel component  $P.q P_\mu /P^2$  of the instant form  translates in the LF form  as  $q_{3\mu} = z P_n n_\mu $, where $P_n = P.{\tilde n} $, and $z = n.q / n.P$. As a check,      
${\hat q}^2 = q_\perp^2 + z^2 M^2$ which shows that $zM$ plays the role of the third component of ${\hat q}$ on LF. Next,  we collect   some of the    
more important   definitions / results of the LF formalism  \cite{Mitr99} 
\begin{eqnarray}
q_\perp & = & q - q_n n ;   {\hat q} = q_\perp + z P_n n;  z  =  q.n / P.n ;   q_n  =  q.{\tilde n} ;    \\  \nonumber    
 P_n  &=  &P. {\tilde n};    P.q  =  P_n q.n + P.n q_n ;  {\hat q}.{\tilde n} = P_\perp. q_\perp = 0 ;  \\  \nonumber 
P.{\hat q} & = & P_n q.n ;  {\hat q}^2 = q_\perp ^2 + M^2 z^2 ; P^2 = - M^2 
\end{eqnarray}   
For a $qqq$ baryon, there are two internal momenta, each separately  satisfying the relations (2.1). Note that for any 4-vector $A$, $A.n$ and $- A_n$ 
correspond to   $1 / \sqrt{2}$ times the usual light front quantities $A_\pm = A_0 \pm A_z$  respectively. But since a $physical$ amplitude must not depend on the orientation $n$,  a simple device termed $Lorentz-completion$ via  the $collinear$ trick \cite{Mitr99} yields a Lorentz-invariant amplitude      
for a transition process  with  $three$ external lines $P, P', P'' (= P + P') $ as follows. Since collinearity implies $P_\perp. P'_\perp =0$,  the 4-scalar  
product $  P.P' = P.n P_n' + P_n P'.n + P_\perp. P'_\perp $ simplifies to $ P.n P_n' + P_n P'.n $. Then   ` Lorentz completion ' simply amounts to 
$reversing$  the last step  via  the `zero' quantity  $P_\perp. P'_\perp $, so  as to recover  the Lorentz invariant quantity $P.P'$ at the end !  And 
as a practical  simplification, one does not even have to use the $n, {\tilde n}$ symbols;
it suffices to use the more familiar light-front components $A_\pm$  for the covariant quantities $ \sqrt{2} [A.n,- A_n] $ respectively.   For ready reference,  the precise correspondence between the  instant and LF  definitions of the  `parallel (z)'  and `time-like (0) ' components of the     
various 4- momenta for a $qqq$ baryon ( i = 1,2,3) \cite{Mita99, MitS01} : 
\begin{equation}\label{2.2}
p_{iz}; p_{i0} = \frac{M p_{i+}}{P_+}; \frac{M p_{i-}}{2 P_-} ; \quad {\hat  p}_i  \equiv  \{ p_{i\perp} , p_{iz}\}
\end{equation}
The last part of Eq.(2.2)  defines a covariant  3-vector on the LF that will frequently  appear in  the reduced 3D BSE for the $qqq$ proton.   
Our goal is to write down  ( and solve) the Master equation for  three fermion quarks complete with all internal d.o.f.'s,  in the presence of both $qq$ and     
direct $qqq$ forces which reads  \cite{CGSM89} : 
\begin{eqnarray}\label{2.3}
\Psi (p_1p_2p_3) &=&  \sum_1^3 S_F(p_1) S_F(p_2) g_s^2 \int \frac{d^4 q_{12}'}{(2 \pi)^4} \gamma_\mu^{(1)} \gamma_\nu^{(2)}
  D_{\mu\nu} (k_{12}) \Psi(p_1', p_2', p_3)  \nonumber  \\ 
                            &  &  +  S_F(p_1) S_F(p_2) S_F(p_3) \int \frac{d^4 q_{12}' d^4 p_3' }{(2 \pi)^8 }  V_{qqq} \Psi(p_1' p_2' p_3')
\end{eqnarray} 
where  the definitions  for the various momenta,  and the phase conventions for the quark propagators  are those of  \cite{CGSM89}, while the   
direct 3-quark interaction $V_{qqq}$ in the last term is $new$, and  given by (1.2).  The central problem is now the reduction of this Master equation (2.3)
 through the three steps (A, B, C)  vide Sect (1.2), as outlined below.     

\subsection{ Reduction of Master Eq from $\Psi$  to $\Phi$ }

Since permutation symmetry plays a crucial role for a $qqq$ hadron for fermion quarks,  we  define  at the outset a  pair of internal variables  
$(\xi; \eta)$  with, say,  the  index $\#3$ as basis,  as \cite{Mita99} 
\begin{equation}\label{2.4}
\sqrt{2} \xi_3 = p_1-p_2 ; \quad \sqrt{6} \eta_3 = -2p_3+p_1+p_2; \quad
P=p_1+p_2+p_3 
\end{equation}
where the  time-like and space- like parts of each are given by (2.2), and the corresponding 3-vector defined as  ${\hat p}_i \equiv \{p_{i\perp}, p_{iz}\}$.  
Two identical sets of momentum pairs $\xi_1, \eta_1$ and $\xi_2, \eta_2$ are similarly defined, but can be expressed in terms of the set (2.4) 
via permutation symmetry.  Now Step A, which is designed   to dispose of the fermion d.o.f.'s  for a $qqq$ system, consists in   
defining  an auxiliary scalar  function $\Phi$ related to the actual BS wave function $\Psi$ by \cite{CGSM89} 
\begin{equation}\label{2.5}
\Psi = \Pi_{123} S_{Fi}^{-1}(-p_i) \Phi (p_i p_2 p_3) W(P)  
\end{equation}
Here the quantity $W(P)$ is independent of the internal momenta but includes the  spin-cum-flavour wave functions $\chi, \phi$ of the 3 quarks 
involved  (see  \cite{MiMi84  for notation and other details} :      
\begin{equation}\label{2.6}
W(P) = [ \chi' \phi' + \chi'' \phi'' ] / \sqrt{2}
\end{equation}
where $\phi'$, $ \phi''$ are the standard  flavour functions of mixed symmetry \cite{Feyn71} [not to be confused with the 3D wave function $\phi$ !], 
and $\chi'$ ,$ \chi''$ are the corresponding relativistic spin functions. The latter may be  defined either in terms of the quark \# indices as in Eqs (1.2)  
or (2.3), or sometimes more conveniently in a common Dirac matrix space as \cite{MiMi84, Blan59} 
\begin{equation}\label{2.7}
|\chi'> ; |\chi''> = [\frac{M - i\gamma.P}{2 M}[i\gamma_5; i{\hat \gamma}_\mu / \sqrt{3}] C / \sqrt{2}]_{\beta\gamma} \otimes 
[[1; \gamma_5{\hat \gamma}_\mu] u(P)]_\alpha
\end{equation} 
where the first factor is the $\beta \gamma$-element of a 4 x 4 matrix in the joint spin space of the  quark \#s 1, 2  \cite{Blan59}, and the second factor      
the $\alpha$ element of a 4 x 1 spinor in the spin space of quark \# 3; $C$ is a charge conjugation matrix with the properties \cite{Davi65}
$$ - {\tilde \gamma}_\mu = C^{-1}\gamma_\mu C ;  {\tilde \gamma}_5 = C^{-1} \gamma_5 C ; $$
and ${\hat \gamma}_\mu$ is the component of $\gamma_\mu$ orthogonal to $P_\mu$. Finally, the representations of the flavour functions $\phi', \phi''$
satisfy the following relations in the "3" basis \cite{MiRo67}  
$$ < \phi'' | 1 ; {\vec \tau}^{(3)} | \phi'' > =  < \phi' | 1 ; - \frac{1}{3} {\vec \tau}^{(3)}  | \phi' > $$ 

\subsection{Gordon Reduction  on $ V_{qq3}$ \& $V_{qqq}$}

At this stage  we indicate  the effect of Gordon reduction on the pairwise kernels $V({\hat \xi}_i{\hat \eta}_i)$ and the 3-body kernel $V_{qqq}$,  following  the original treatment of  \cite{Mitr81}, (also quoted in [8] ).  The strategy lies in a close scrutiny  of Eqs.(2.3-2.5) with a view to  eliminate  
the Dirac matrices in favour of the  Pauli matrices $\sigma_{\mu\nu}$. To that end, we express  the Dirac propagators $S_F(p_i)$ as 
$$ i S_F (p_i)  = (m_q - i \gamma^{(i)}. p_i) /  \Delta_i ; \quad \Delta_i = m_q^2 + p_i^2  $$ 
Now employing the notation 
\begin{equation}\label{2.8}
   V^{(i)}_\mu =  (m_q - i \gamma^{(i)}. p_i) i\gamma^{(i)}_\mu 
\end{equation} 
the result of Gordon reduction is expressed by the formula \cite{Mitr81, CGSM89} 
\begin{equation}\label{2.9}
 V^{(1)}. V^{(2)}=  - M_{12}^2 - ({\hat q} + {\hat q}')^2 + i ({\hat q} + {\hat q}'). ({\hat \sigma}_1+ {\hat \sigma}_2) \times ({\hat q} - {\hat q}') \nonumber \\  
- \sigma^{(1)}_{ij} \sigma^{(2)}_{ik} (q-q')_j (q-q')_k    
\end{equation} 
where we have used a mixed vector-tensor notation for the various 3-vectors on the light-front in the sense of (2.2) and a short-hand notation $q$ 
for $q_{12}$.   It is now possible to identify  the pairwise kernel arising from the first group of terms on the rhs of  Eq.(2. 3)  after expressing it  
in   LF notation of  Eq.(2.2)  as follows:
\begin{equation}\label{2.10}
V_{qq3} \equiv  V({\hat \xi}_3,{\hat \xi}_3'') = g_s^2  V^{(1)}. V^{(2)} \times \frac{-4}{3 } \times \frac{1}{ ( {\hat q}_{12} - {\hat q}''_{12})^2}
\end{equation} 
where we have included  a net color factor of $(-4 /3)$  arising from a folding of the  factor $F_{12}$  for each $qq$ pair in a $3^*$ state, into 
$\lambda_3 / 2$ for the spectator quark \# 3 in a $3$ state;  the corresponding gluon propagator 
has been taken on the light-front, in the notation of (2.2).  Similarly the 3-body kernel $V_{qqq}$ in Eq.(2.3) may be identified as a sum of 
3 terms arising from a folding of (1.2) with the 3 Dirac factors $ (m_q - i \gamma^{(i)}. p_i)$   in the second term of (2.3).  
\begin{equation}\label{2.11}
 V_{qqq} = \sum_{123} (-4 / 3) g_s^4 [ V^{(1)}. {({\hat k}_2 - {\hat k}_3)} \times  V^{(2)}. V^{(3)}   + \{2\} + \{3\}]/ \{ {\hat k}_1^2 {\hat k}_2^2 {\hat k}_3^2 \}          
 \end{equation} 
using the notation $V^{(i)}_\mu $ of (2.8). The net effect of the (antisymmetric) combination of the color matrices $\lambda$  is represented by the   
factor $(-4 / 3)$ sitting in front; and  $k_i = p_i - p_i'$. The `bosonized' form of Eq. (2.3) is now 
\begin{eqnarray}\label{2. 12}
\Phi (p_1p_2p_3) &=& \sum_{123} \frac{ 1 }{ \Delta_1 \Delta_2}  \int \frac{d^4 q_{12}'}{(2 \pi)^4}V({\hat \xi}_3,{\hat \xi}_3'') \Phi(p_1', p_2', p_3)  \nonumber  \\ 
& +&  \frac{1}{\Delta_1 \Delta_2 \Delta_3} \int \frac{d^4 q_{12}' d^4 p_3' }{(2 \pi)^8 }  V_{qqq} \Phi(p_1' p_2' p_3')
\end{eqnarray} 
in terms of  the structures  (2.10) and (2.11) for the 2- and 3-body kernels respectively.  And the Gordon reduction formula  (2.9 ) may be repeatedly used to used to  further simplify the structures  of  $V_{qq}$[Eq (2.10)]  and $V_{qqq}$[ Eq. (2.12)], which are collected as follows.  

\subsubsection{ Further Reduction  of  $V_{qq3}$ \& $V_{qqq}$ }

\begin{figure}
\center{\includegraphics[width=11cm]{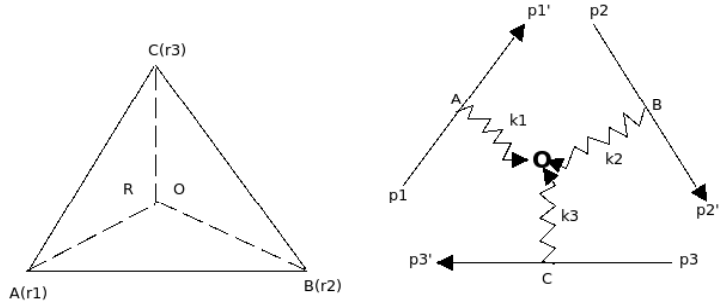}}
\caption{(a) Pictorial view of 2- \& 3- body interactions; (b) `Mercedez-Benz' diagram for qqq-force } 
\end{figure}

The  2-body and 3-body interactions  are pictorially represented by fig 1(a) where  the vertices of the triangle $ABC$  stand for the position 
vectors ${\bf ( r_1, r_2, r_3) }$ of the 3 quarks  in configuration space, while the centroid $O$ has the position vector 
$ {\bf R} = {\bf ( ( r_1+ r_2+ r_3)} /3$.  The corresponding  momenta of the gluon lines connecting them for pairwise $qq$ interactions are precisely the 
quantities  $q_{ij}- q'_{ij}$ that appear as arguments in Eq. (2.10), while the momenta associated with the gluon lines connecting all the 3 quarks 
together, a la the $Y$-shaped fig.(1b), are the quantities ${\hat k}_i$ corresponding to the vectors ${\bf (r_i - R)}$. In the notation of Eq.(2.4),  the former 
momenta are  precisely ${\hat \xi}_i$,  while the latter are the $dual$ variables  $ {\hat \eta}_i$. [This interpretation will prove useful in the subsequent analysis]. 
\par 
The pairwise terms $V_{qqi}$ were earlier  considered in  detail  [8]. The `convective'  terms are of the form $(p_1 + p_2).(p_1' + p_2')$
which simplifies  to $ - M_{12}^2 \approx -4 M^2 / 9$, plus smaller terms. The spin-orbit terms are unimportant for the ground state of the  
$qqq$ proton, but  the spin-spin terms survive  for its ground state $S = 1/2$; their angular average works out as : 
$$ \frac{ 2}{3}\sum _{ij}{\hat \sigma}_i.{\hat \sigma}_j   = \frac{1}{3}[ \Sigma^2 - 9] $$ 
where we have introduced the total spin operator $\Sigma$ for the proton state, with $\Sigma^2 =3$. Collecting these results the sum 
of the 3 pairwise interactions (2.10) work out as
\begin{equation}\label{2.13}
V_{qq3} = g_s^2 \frac{8}{27}  \times [ \sum \frac{4 M^2 p_{3z} }{({\hat q}_{12} - {\hat q}''_{12})^2} - 3 p_{3z}(\Sigma^2 -9)] 
\end{equation}
We now  consider the more interesting term $V_{qqq}$, Eq.(2.12).   
The numerator can be broken up into 3 distinct groups : a) convective; b) spin-orbit; c) spin-sp[in-spin. Using 
the notation $2{\bar p}_i = p_i + p'_i$, the convective part  vanishes (due to antisymmetry with too few d.o.f.'s) : 
$$ \sum 4 [{\bar p}_1. {\bar p}_2] [2 {\bar p}_3. (k_1 - k_2)]  \Rightarrow 0 $$ 
The spin-orbit part comes from two groups of terms, one of which vanishes: 
$$ \sum 2 {\bar p}_1. (k_1 - k_2) [ {\hat \sigma}_3. ({\bar {\hat p}}_3 \times {\hat k}_3) \Rightarrow 0 $$ 
where a mixed notation is used for the  Pauli matrices, light-faced tensor for the 4D and hatted vector for the 3D forms.   
The other group gives rise  to a non-zero contribution as
\begin{equation}\label{2.14}
 \sum 4 {\bar p}_1.{\bar p}_2 [ i\sigma^{(3)}_{\mu\nu} k_{3\nu}(k_{1\mu} - k_{2\mu})] \approx (-4i M^2 / 9)
\sum_i {\hat \sigma}_i . {\hat \xi}_3\times {\hat \eta}_3
\end{equation} 
in the notation of  Eq.(2.4). Note the overall antisymmetry of the last factor (independent of the index 3).    
\par 
Finally  the (totally antisymmetric) spin-spin-spin contribution $SSS$ is defined  as : 
\begin{equation}\label{2.15}
 SSS  =  \sum_{123}[ i\sigma^{(1)}_{\mu\nu}{\hat k}_{1\nu}({\hat k}_{2\mu} -{\hat k}_{3\mu}) i\sigma^{(2)}_{\lambda\rho}{\hat k}_{2\rho}
 i\sigma^{(3)}_{\lambda\sigma}{\hat k}_{3\sigma}
\end{equation}      
The simplification of this expression (which is elementary) is helped by the identity $\sigma^{(i)}_{\mu\nu} = \epsilon_{\mu\nu a} {\hat \sigma}^{(i)}_a$. 
The next step is angular averaging over each of the independent 3-vectors involved in each of the three terms above, namely,  
$<{\hat k}_i {\hat k}_j> = {\hat k}^2 \delta_{ij} / 3 $. The resulting expression for $SSS$ is
$$ SSS = \frac{2i}{9 } \{ {\hat k}_1^2 {\hat k}_2^2 {\hat k}_3^2 \} {\hat \sigma}_1. {\hat \sigma}_2 \times {\hat \sigma}_3 \sum_i \frac {1}{{\hat k}_i^2} $$
Substituting for $SSS$ in Eq (2.11) and ignoring  the spin-orbit terms (for the ground state of the proton) the net 3-body force is 
\begin{equation}\label{2.16}
V_{qqq} = \frac{-8i g_s^4}{27} ( {\hat \sigma}_1. {\hat \sigma}_2 \times {\hat \sigma}_3)  \sum_i \frac {1}{{\hat k}_i^2}  
\end{equation}

\section{  3D Reduction By  Green's Function  Method} 

\setcounter{equation}{0}
\renewcommand{\theequation}{3.\arabic{equation}}

We are now in a position to implement Step C of our program ( establishing a 3D-4D interconnection  between the corresponding wave functions) , 
for which it is  convenient to employ the Green's function approach \cite{Mita99}.   Calling the 4D Green's functions associated with 
$\Psi$ and $\Phi$ by $G_F$ and $G_S$ respectively, the  connection between them analogously to Eq.(2.5) may be written as 
\begin{equation}\label{3.1}
G_F (\xi\eta;\xi'\eta')=  W(P) \otimes  { \Pi_{123} S_{Fi}^{-1}(-p_i)} G_S (\xi\eta;\xi'\eta') {\Pi_{123} S_{Fi}^{-1}(-p_i')} {\bar W}(P') 
\end{equation} 
where we have indicated the 4-momentum arguments of the Green's functions involved,  in a common $S_3$ basis $(\xi, \eta)$,  and expressed the 
spin-flavour dependence of $G_F$ as a  matrix product  implied by the notation $W (P) \otimes {\bar W} (P')$.  
\par
Now to show the 3D-4D interconnection it is enough to work at the level of the `scalar' Green's function $G_S$ ( relabelled  as  $G$ for simplicity), 
since it is trivial to include the spin-flavour d.o.f.'s in a matrix notation  via (2.5) later.  The main steps for the scalar $qqq$ Green's function are 
shown next, while Appendix A  sketches the corresponding process for  a typical $qq$ subsystem, from 3D reduction to 4D reconstruction, so as to serve as a simpler prototype for the actual $qqq$ case.

\subsection{ Reduction from 4D $G(\xi \eta; \xi' \eta' )$ to 3D ${\hat G}({\hat \xi}{\hat \eta};{\hat \xi}'{\hat \eta}')$ }  

We shall now outline a Green's function method to establish a 3D-4D interlinkage between the 4D $\Phi$ and the 3D $\phi$, following a generalization  
of the procedure developed some time ago \cite{Mita99}, so as to include the effect of the direct 3-body force $V_{qqq}$. To that end we shall rederive     the $qq$ denominator functions $D_{ij}$ of \cite{Mita99} arising from single integrations over $d\xi_{i0}$, to show that they are $exactly$ proportional to the  3D  denominator function $D_{123}$  arising from a double integration over $d\xi_0 d\eta_0$ in  the direct 3-body  term  to be considered here. [This is a  big improvement over the earlier derivation [22] where this property was missing,  and helps   pave the way for putting together the effects of both 2- and 3-body forces within a common dynamical  framework ].       
\par
The full 4D Green's function for `scalar' quarks is $G(\xi\eta;\xi'\eta')$ (taking out the 
c.m $\delta$-fn), while its  3D counterpart  ${\hat G}$ is \cite{Mita99}
\begin{equation} \label{3.2}
{\hat G}({\hat \xi}{\hat \eta};{\hat \xi}'{\hat \eta}') =
\int {d\xi_0 d\eta_0 d\xi_0' d\eta_0'} G(\xi\eta;\xi'\eta')
\end{equation}
where the time-like subscript `0'  should be read as the corresponding LF component  in the sense of Eq.(2.2). Note that 
both $G$ and ${\hat G}$ are $S_3$ symmetric, since the combination $d\xi_0 d\eta_0$  has this property.  But the two hybrid Green's functions   
defined as \cite{Mita99}
\begin{eqnarray}\label{3.3}   
{\tilde G}_{3\xi}({\hat \xi}_3 \eta_3;{\hat \xi}'_3 \eta'_3) &=& \int {d\xi_{30} d\xi_{30}'} G(\xi\eta;\xi'\eta'); \nonumber \\
{\tilde G}_{3\eta}(\xi_3{\hat \eta}_3;\xi_3'{\hat \eta}_3') &=& \int {d\eta_{30} d\eta_{30}'} G(\xi\eta;\xi'\eta') 
\end{eqnarray}
are $not$ $S_3$ symmetric  (hence they are indexed), since the integration now involves only $one$ 
of the two $\xi$,$\eta$ variables. Next, the 4D  $G(\xi\eta; \xi'\eta')$ satisfies a BSE  of the form   
\begin{eqnarray}\label{3.4}
i(2\pi)^4 G(\xi\eta;\xi'\eta') &=& \sum_{123} \int \frac{ d^4\xi''}
{4\Delta_1\Delta_2} V({\hat \xi}_3,{\hat \xi}_3'' )  G(\xi_3''\eta_3;\xi_3'\eta_3')  \nonumber \\ 
                                          &+& \int \frac{ d^4 \xi'' d^4 \eta''}{9 (2\pi)^4 \Delta_1\Delta_2 \Delta_3} V_{qqq} G(\xi_3''\eta_3;\xi_3'\eta_3') 
\end{eqnarray}
which is analogous to Eq. (2.12) for the corresponding wave functions, except for a change of variables wherein  the factors $1/4$  and $1/9$ in the first and second groups  of terms stem from the relations (2.4) among the corresponding variables.  It is understood that both these types of interaction have covariant 3D support, and they  incorporate the effect of `Gordon reduction'  as outlined in Sect 2.3 above.    The 3D reduction of (3.4) is now achieved by integrating it  a la  (3.2), leading to a structure of the form :  
\begin{eqnarray}\label{3.5}
(2\pi)^3 {\hat G}({\hat \xi}{\hat \eta};{\hat \xi}'{\hat \eta}') &=& \sum_{123} \frac{1}{2\sqrt{2} D_{12}} \int d^3{\hat \xi}_3'' V({\hat \xi}_3,{\hat \xi}_3'')  {\hat G}({\hat \xi}_3''{\hat \eta}_3'';{\hat \xi}_3'{\hat \eta}_3')  \nonumber \\
&+ & \frac{1 }{3\sqrt{3} (2\pi)^3 D_{123} }\int d^3 \xi'' d^3 \eta'' V_{qqq} {\hat G}(\xi_3'' \eta_3'' ; \xi_3' \eta_3') 
\end{eqnarray}
To explain the structure of certain factors, Eq (3.5)  shows that we have two types of  3D denominator functions $D_{ij}$ and $D_{123}$,  associated with  pairwise 2-body and direct 3-body forces  respectively. It is easily shown that they are simply related to each other.  To that end, Appendix A  already shows the structure of $D_{ij}$ :   
\begin{equation}\label{3.6}
D_{12} = \frac{M D_{12+}}{P_+}; \quad   D_{12+} = 2M [\omega_{1\perp}^2 p_{2+}+ \omega_{2\perp}^2 p_{1+}- P_{12-}p_{1+}p_{2+}] 
\end{equation}
where $ \omega_{i\perp}^2 $ equals  $m_q^2 + p_{i\perp}^2$.  Using the  on-shellness ($\Delta_3 =0$) of the spectator ($\#3$)  then gives
$ P_{12-}=P_- - p_{3-}$ which reduces further to  $P_- - \omega_{3\perp}^2 / p_{3+}$. Its  substitution back in (3.6) leads  to the $S_3$ symmetric result:
\begin{equation}\label{3.7}
p_{3+} D_{12+} \equiv \frac{P_+^2}{2M^2} D_{123} = 2\sum_{123} \{p_{2+} p_{3+}\omega_{1\perp}^2\} -2p_{1+} p_{2+} p_{3+} P_- 
\end{equation}
Here we have anticipated the structure of  $D_{123}$ associated with the $V_{qqq}$ term whose formal definition is 
\begin{equation}\label{3.8}
\frac{1}{D_{123 }} =  \int \frac { P_+^2 dq_{12-} dp_{3-}}{ 4 M^2 (2i \pi)^2 \Delta_1 \Delta_2 \Delta_3} 
\end{equation}
where the (double) contour integrations in the indicated variables leads to the desired result.   The resultant 3D BSE for ${\hat G}$ is now      
\begin{eqnarray}\label{3.9}
(2\pi)^3 D_{123} {\hat G}({\hat \xi}{\hat \eta};{\hat \xi}'{\hat \eta}') &=& \sum_{123} \frac{p_{3z}}{\sqrt{2}}  \int d^3{\hat \xi}_3'' V({\hat \xi}_3,{\hat \xi}_3'')  {\hat G}({\hat \xi}_3''{\hat \eta}_3'';{\hat \xi}_3'{\hat \eta}_3')  \nonumber \\
&+& \frac{1}{3\sqrt{3} (2\pi)^3} \int d^3 \xi'' d^3 \eta'' V_{qqq} {\hat G}(\xi_3'' \eta_3'' ; \xi_3' \eta_3')
\end{eqnarray}
The structure of this equation reveals some interesting symmetries, when (2.13) and (2.16) are substituted for $V_{qq3}$ and $V_{qqq}$ respectively. 
Namely, in the first group of terms the $\eta$ variable is the spectator since the integration is only over the $\xi$ variable, while  in the second group of terms, their relative roles are interchanged with  the $\xi$ variable  effectively a spectator ! This fact is not immediately apparent because of the double integration involved, but a little reflection shows that  the absence of the $\xi$ variable in (2.16) effectively ensures its spectator status in the argument 
of the corresponding Green's function. We shall see this feature more clearly in the coordinate space representation of the $\phi$ equation 
(corresponding to Eq (3.9) for ${\hat G}$, to be considered in Section 5. But first we turn our attention to the $reconstruction$ of the 4D Green's  function 
from the 3D form ${\hat G}$ satisfying  (3.9).  To that end  we note that the $qq$ and $qqq$ group of terms in (3.9) need different strategies and are best handled separately, one at a time,  (temporarily) ignoring the presence of the other. 
 
\section { Reconstruction of 4D Wave Function}

\subsection{Reconstruction of $G$ with $qq$ Forces Only}

\setcounter{equation}{0}
\renewcommand{\theequation}{4.\arabic{equation}}

Considering first the $qq$ group, we  proceed exactly as in (A.10) of Appendix A for the $qq$ system. Namely,  first express    ${\tilde G}_{3\eta}$, eq.(3.3), in terms  of the 3D ${\hat G}$, using the notation of (2.2) and the result of (3.7): 
\begin{equation}\label{4.1}
{\tilde G}_{3\eta}(\xi_3{\hat \eta}_3; \xi_3'{\hat \eta}_3')
=\frac{D_{123}}{2i\pi p_{3z}\Delta_1 \Delta_2}   \\
{\hat G}({\hat \xi}{\hat \eta};{\hat \xi}'{\hat \eta}')
 \frac{ D'_{123}}{2i\pi p_{3z}'\Delta_1' \Delta_2'} 
\end{equation}
 In a similar way the fully 4D $G$ function is expressible in terms
of the hybrid function ${\tilde G}_{3\xi}$ as
\begin{equation}\label{4.2}
G(\xi\eta;\xi'\eta') = \sum_{123} \frac{D_{123}}{2i\pi p_{3z}\Delta_1 \Delta_2} 
{\tilde G}_{3\xi}({\hat \xi}_3\eta_3; {\hat \xi}_3' \eta_3')             
\frac{ D'_{123}}{2i\pi p_{3z}'\Delta_1' \Delta_2'}
\end{equation}
Now since the ${\tilde G}_{3\xi}$ function is not determined from 
$qqq$ dynamics alone, we  invoke an ansatz similar to, but more symmetrical than,  \cite{Mita99} :
\begin{equation}\label{4.3}
{\tilde G}_{3\xi}({\hat \xi}_3 \eta_3; {\hat \xi}_3'  \eta_3')
 = {\hat G}({\hat \xi}{\hat \eta}; {\hat \xi}' {\hat \eta}') F(p_3, p_3')
 \end{equation}
where the balance of the $p_3$ (spectator) dependence is in the (as yet unknown) 
$F$ function, subject to an explicit self-consistency check. To that end,
try the ( symmetrical)  Lorentz-invariant form  
\begin{equation}\label{4.4}
F(p_3, p_3') = \frac{A_3}{\Delta_3} \delta(\Delta_3 -\Delta_3')   
\end{equation}
and integrate both sides of (4.4) w.r.t. $dp_{30} dp_{30}'$, in the notation of (2.2), to show
that the consistency check is met with  the mass shell value of the spectator momentum :  
$$ A_3 = \frac{4p_{3z} p_{3z}'}{2i\pi }$$  
Substitution from $A_3$ in (4.4) then gives the symmetrical  form   
$$  
F(p_3,p_3') =  4p_{3z}p_{3z}' \frac{\delta(\Delta_3 - \Delta_3')}{2\pi i \Delta_3}
$$
whence  the 4D $G$-fn  in terms of ${\hat G}$ via the
sequence (4.3) and (4.2), works out as
\begin{equation}\label{4.5}
G(\xi\eta;\xi'\eta') = \sum_{123} \frac{D_{123}}{\Delta_1 \Delta_2} 
{\hat G}({\hat \xi}{\hat \eta};{\hat \xi}'{\hat \eta}')             
\frac{ D'_{123}}{\Delta_1' \Delta_2'} \times \frac{\delta(\Delta_3 - \Delta_3')}{(2\pi i)^5 \sqrt{\Delta_3 \Delta_3'}}
\end{equation}
 
\subsection{ Reconstruction of $G$ with $qqq$ Forces Only}

Next we consider only the $qqq$ group of terms  for a reconstruction from ${\hat G}$  ( Eq (3.5))   to $G$ (Eq (3.4)). This case is more akin to 
the $qq$ case of Appendix A  in the sense that if we use the collective indices $(\xi, \eta) \equiv \rho$ and $(\xi', \eta') \equiv \rho'$ for the initial and final state arguments of the $G$-function we can  define two kinds of  ${\tilde G}$ as follows.  
\begin{equation}\label{4.6}
{\tilde G}({\hat \rho}; \rho') = \int d\rho_0 G(\rho; \rho'); \quad 
{\tilde G}(\rho; {\hat \rho}') = \int d\rho_0' G(\rho; \rho')
\end{equation}      
where the integrals on the RHS are each of the double integral type, so that  the definitions of the denominator functions involved 
are given by Eq.(3.8). Thus, following (A. 9),  ${\tilde G}$ and ${\hat G}$ are connected as follows 
\begin{equation}\label{4.7}
{\tilde G}(\rho ,{\hat \rho}') = \frac{D_{123}({\hat \rho})}{(2i\pi)^2 \Delta_1 \Delta_2 \Delta_3}
{\hat G}({\hat \rho};{\hat \rho}')
\end{equation}    
together with a second one with the roles of $\rho, \rho'$ interchanged.  Continuing exactly as in Appendix A, $G$ of Eq (3.4) gets expressed in terms of 
${\tilde G}$  on the RHS, since  the interaction $V_{qqq}$  does $not$ involve  the variables $\rho_0'$. Thence another 
application of Eq.(4.7) for expressing ${\tilde G}(\rho; {\hat \rho}')$  in terms of ${\hat G}$ finally yields the desired connection:
\begin{equation}\label{4.8}
G(\rho; \rho') = \frac{D_{123}({\hat \rho})}{(2i\pi)^2 \Delta_1 \Delta_2 \Delta_3} 
{\hat G}({\hat \rho};{\hat \rho}') \frac{D_{123}({\hat \rho}')}{(2i\pi)^2 \Delta_1' \Delta_2' \Delta_3'}
\end{equation}
where the symbol $\rho$ stands collectively for $(\xi, \eta)$, as does $\rho'$. Note that in this case the reconstruction of $G$ in terms of 
of (the fully reduced) ${\hat G}$ is unique, and does not require any extra ansatz like (4.4). 

\subsection{ Putting Both $V_{qq}$ \& $V_{qqq}$ Together} 
 
We have now two distinct types of 3D-4D interconnections valid for pure $qq$ and pure $qqq$ forces respectively. But when both types of forces 
are  present, an exact  interconnection is very difficult to derive. We shall therefore strive for an approximate but sufficiently realistic solution 
based on the relative strengths of the two forces, namely,  an overwhelming preponderance of $qq$ over $qqq$ forces. To  give effect to this 
strategy, we first note that  the 3D level does $not$ involve any  approximation since  Eq. (3.9) treats both types of interaction on par.  
 It is only at the 4D level of reconstruction that an approximation is necessary for putting the 
two forces together. Now  taking account of the dominance of $qq$ over $qqq$ forces, the simplest choice is to prefer the structure of  Eq (4.5)   
over that of Eq. (4.8).  And although the ratio of the two 3D-4D interconversion jackets in (4.5) over (4.8)  is a singular quantity 
 $$   \delta(\Delta_3 - \Delta_3') \times (2\pi i) \sqrt{\Delta_3 \Delta_3'} $$  
it causes  no harm for physical amplitudes since  such singular quantities  get ironed out on  integration over the various momenta \cite{MitS01}. 
With this understanding, Eq.(4.5) represents the  reconstruction of the full 4D Green's function $G(\xi\eta;\xi'\eta') $ in terms of the 3D quantity 
${\hat G}({\hat \xi}{\hat \eta};{\hat \xi}'{\hat \eta}')$ where the latter now satisfies Eq (3.9) which includes the effect of both $qq$ and $qqq$ forces. 

\subsection{3D-4D Interconnection for  Wave Functions}

 Finally the  spectral representations of $G(\xi\eta;\xi'\eta') $ and ${\hat G}({\hat \xi}{\hat \eta};{\hat \xi}'{\hat \eta}')$  for the $qqq$ system, exactly  
on the lines of (A.12-A.13) for a $qq$ subsystem  near a bound state pole $P^2 = -M^2$ , are  
\begin{equation}\label{4.9}
G(\xi\eta; \xi'\eta') = \sum_n \Phi_n(\xi,\eta) \Phi_n^* (\xi',\eta')/ (P^2 +M^2);
\end{equation}
and a similar equation for ${\hat G}$ vis-a-vis $\phi$. These  give the connection between the  4D wave function $\Phi$ which 
satisfies  Eq (2.12),  and the 3D wave function $\phi$ corresponding to the  Green's function  ${\hat G}$ which satisfies (3.9).  For purposes of 
evaluating transition amplitudes via Feynman diagrams, it is convenient to index $\Phi$ as  $ \Phi_1 + \Phi_2 + \Phi_3$,   and a corresponding indexing for the associated vertices as $ V = V_1 + V_2 +V_3$ , as in ref [22],  so as to keep track of which quark is involved in which vertex.       
\begin{equation}\label{4.10}
\Phi_3 (\xi, \eta)  \equiv \frac{V_3}{\Delta_1 \Delta_2 \Delta_3} = \frac{D_{123}}{\Delta_1 \Delta_2} \phi({\hat \xi}, {\hat \eta})              
 \frac{\sqrt{\delta(\Delta_3 )}}{(2\pi i)^{5/2} \sqrt{\Delta_3 }}  
\end{equation}
The $\delta$-function in eq.(4.7) has nothing to do with any connectedness  problem;  
see ref.\cite{Mita99} for detailed reasons.     
Finally, the use of Eq.(2.5) with (4.10) yields an explicit structure for the actual (fermionic) wave function $\Psi$ as 
\begin{equation}\label{4.11}
\Psi (\xi, \eta)  = \Pi_{123} S_F(p_i) D_{123}\sum_{123} [ \phi({\hat \xi}, {\hat \eta})              
\frac{ \sqrt{\delta(\Delta_3 ) \Delta_3}}{(2\pi i)^{5/2}}] \times W(P) 
\end{equation}  

\section{ Complete $\phi$ Equation  In Coordinate Space } 

\setcounter{equation}{0}
\renewcommand{\theequation}{5.\arabic{equation}}

Our final task is to set up (and solve)  the 3D BSE for $\phi$ which  may be inferred from (3.9) by making use of a spectral representation similar 
to (4.9), and going to the pole $P^2 = - M^2$ : 
\begin{eqnarray}\label{5.1}
(2\pi)^3 D_{123} \phi({\hat \xi}, {\hat \eta})  &=& \sum_{123} \frac{p_{3z}}{\sqrt{2}}  \int d^3{\hat \xi}_3'' V_{qq3} \phi({\hat \xi}_3'', {\hat \eta}_3)   \nonumber \\
&+& \frac{1}{3\sqrt{3} (2\pi)^3} \int d^3 \xi'' d^3 \eta'' V_{qqq} \phi({\hat \xi}_3'', {\hat \eta}_3'')
\end{eqnarray}
where $V_{qq3}$ is given by (2.13);  $V_{qqq}$  by (2.16); and $D_{123}$ by (3.7).  To transform this equation in coordinate space, define the 
combinations analogous to (2.4) as 
\begin{equation}\label{5.2}
\sqrt{2} s_3 = r_1 - r_2 ; \quad \sqrt{6} t_3 = -2 r_3 + r_1 + r_2 
\end{equation} 
Then a Fourier transform to the $coordinate-space$ representation gives for the pairwise terms  $V_{qq3}$ the coulombic  structure   
\begin{equation}\label{5.3}
V^{(2)}(s) =   \int \frac{d^3 {\hat \xi}_3'}{ (2\pi)^3}\exp{[i{\hat \xi}_3'. s_3]}  \nonumber \\
 \times [ \sum  \frac{8 M^2 g_s^2 p_{3z}}{27 ({\hat \xi}_3- {\hat \xi}'_3)^2}  - \frac{  g_s^2}{9} ( \Sigma^2 -9) p_{3z} ]  
\end{equation}
which  multiplies the coordinate space wave function $\phi(s, t)$ ( $S_3$ symmetric in its arguments). 
Similarly, the ($S_3$ symmetric) double Fourier transform of  $V_{qqq}$ is  defined as 
\begin{equation}\label{5.4}
V^{(3)} (s, t) = \frac{-4i g_s^4}{9} \int \frac{d^3{\hat \xi}' d^3{\hat \eta}'}{3 \sqrt{3}(2\pi)^6}\exp{ [ i{\hat \xi}'.s + i{\hat \eta}'. t]} \nonumber \\
 \times \sum_{123}  \frac{1}{({\hat \eta}_3- {\hat \eta}'_3)^2} [{\hat \sigma}_1. ({\hat \sigma}_2 \times {\hat \sigma} _3)]        
\end{equation}
The  integration  in (5.3) involves a 3D $\delta$-function $\delta^3({\hat s}_3)$ ; and the double integration in (5.4)  additionally  involves  the 
factot  $ 1 / | t_3 | $ due to the $\eta$ integration. The $\delta$ function being  singular may be   `regularized'  by  using a differential representation  (acting on the scalar function $\phi(s,t)$),   with $|{\hat s}_3|$ = $|s_3|$ : 
\begin{equation}\label{5.5}
 \frac{\delta^3 ({\hat s}_3)}{(2 \pi)^3)} \phi(s,t)   =  \Rightarrow \frac{ -1}{ 4\pi |s_3| } [ \partial_{s3}^2 ] \phi(s,t)    
\end{equation} 
The result of integration in (5.3) is then expressed as 
\begin{equation}\label{5.6}
V^{(2)}(s) =    \sum [ \frac{8 M^2 \alpha_s p_{3z}}{27 | s_3| }  + \frac{  \alpha_s ( \Sigma^2 -9) }{ 27 | s_3|}  p_{3z} \partial_{s_3}^2 ]  
\end{equation}
Next  the 3-body term  term (5.4)  integrates, with the help of (5.5),   to 
$$ 4 \alpha_s^2 \frac{[i{\hat \sigma}_1. ({\hat \sigma}_2 \times {\hat \sigma} _3)]}{27 \sqrt{3}} \sum \frac{1}{s_3 t_3}\partial_{s3}^2 $$  
Now  the product  $1 / s_3 t_3$  can be approximately replaced by the  $S_3$ symmetric expression $2/[s_3^2 + t_3^2]$ which can be taken out of the 
summation sign (with index `3' dropped)  to give $\sum \partial_{s3}^2$ which in turn equals $ (3/2) (\partial_s^2 + \partial_t^2) $,  an $S_3$ symmetric sum with index `3' dropped again.   This  finally  leads from (5.4) to the  $S_3$ symmetric  form of the 3-body term in coordinate space: 
\begin{equation}\label{5.7}
V^{(3)}(s,t) = 4  \alpha_s^2 \frac{[i{\hat \sigma}_1. ({\hat \sigma}_2 \times {\hat \sigma} _3)]}{ 9 \sqrt{3}(s^2 + t^2)} [ \partial_s^2 + \partial_t^2 ]  
\end{equation}
For further manipulations  it is convenient to replace the antisymmetric  spin operator $A$ = $ [i{\hat \sigma}_1. ({\hat \sigma}_2 \times {\hat \sigma} _3)]$ 
by one of its possible eigenvalues as follows: Squaring this quantity yields in a simple way 
$$ A^2 =  - A  - 15 + \Sigma^2 ;  \quad \Sigma \equiv  {\hat \sigma}_1+ {\hat \sigma}_2 + {\hat \sigma}_3 $$
from which one obtains the successive results 
$$ A^3 + A^2 +15 A = \Sigma^2 A ;  \quad  A^2 (A +1)^2 = [\Sigma^2 -15]^2 $$
The last equation yields the four solutions 
\begin{equation}\label{5.8}
A = - \frac{1}{2} \pm \sqrt{15 - \Sigma^2 - 1/4} ; \quad A = - \frac{1}{2} \pm \sqrt{\Sigma^2 - 15 - 1/4} 
\end{equation} 
whose substitution in Eqs (5.4) and (5.7)  summarises the full  content  of  the total spin effect of  the $V_{qqq}$ term.  Finally,  the denominator term $D_{123}$ is a differential operator in coordinate space :
\begin{equation}\label{5.9}
D_{123} /4 = \sum_{123}[ (- m_q^2 + \partial_{3\perp}^2)  \partial_{1z}\partial_{2z}]  - i M \partial_{1z} \partial_{2z} \partial_{3z} 
\end{equation}
Thus $\phi$ satisfies the following equation in coordinate space :
\begin{equation}\label{5.10}
D_{123} \phi(s,t) = [V^{(2)} (s) + V^{(3)} (s,t)] \phi(s,t) 
\end{equation} 
where the various operators are given by (5.6) - (5.9).   

\subsection{ Simplified Form  of the $\phi$ Equation (5.10)}

Since  this paper is  intended as a preliminary mathematical framework for the dynamical effect of the spin terms, especially the  spin-rich   $V_{qqq}$ term on the spatial  structure of the proton wave function,  pending  a full-fledged study  of the proton spin anomaly,   we shall  at this stage merely  
 outline a  qualitative procedure to determine the nature  of the solution of  Eq. (5.10),  with particular reference to the possible role of  the eigenvalues (5.8)  of $A$, on the solution of this differential equation. To that end, we shall make some drastic simplifications, starting with the differential operator (5.9) whose principal terms  (up to second order ),  in momentum space,  work out as  
$$ D_{123} \approx  \frac{4 M^2}{9}( \xi_\perp^2 + \eta_\perp^2 ) + [\frac{2M^2}{3} - 2 m_q^2 ] (\xi_z^2 + \eta_z^2) + \frac{4 M^2}{3} (m_q^2 - M^2 /9) $$ 
Further simplification arises with the `special' value $ M = 3 m_q$ which then yields a  simple yet transparent  expression in coordinate space :
\begin{equation}\label{5.11}
D_{123} \approx \frac{4 M^2}{9} [ - \partial_s^2 - \partial_t^2 ]
\end{equation}
an operator with full rotational  invariance and $S_3$ symmetry  in the 3D space ( on light-front)   defined by Eq. (2.2).    
In a similar vein, the 2-body terms (5.6) can be simplified by the replacements $ p_{iz} \approx M / 3 $, so as to yield 
\begin{equation}\label{5.12}
V^{(2)}(s,t) =  \frac{ 8\sqrt{2} M^3 \alpha_s}{27\sqrt{ s^2 + t^2}} + \frac{ M \alpha_s ( \Sigma^2 -9) \sqrt{2}}{ 54 \sqrt{s^2 +t^2} } [\partial_s^2 + \partial_t^2] 
\end{equation} 
where we have made a further simplification based on certain standard inequalities [31] 
$$ \sum_{123} \frac{1}{ | s_3 |} \approx \frac{3 \sqrt{2}}{\sqrt {s^2 + t^2}} $$
And  the 3-body term (5.7) may be written more compactly in terms of of the operator $A$  with eigenvalues (5.8) as 
\begin{equation}\label{5.13}
V^{(3)}(s,t) =   \alpha_s^2 \frac{4 A}{ 9 \sqrt{3}(s^2 + t^2)} [ \partial_s^2 + \partial_t^2 ]  
\end{equation}
One has now a differential equation for (5.10), with the  simplified operators (5.11-13) which exhibit a 6D symmetry. Taking $R^2 = s^2 + t^2$, 
the 6D Laplacian for the ground state of the proton (with all angular d.o.f.'s dropped) takes the simpler  form 
$$ [ 4 M^2 / 9  + \frac{M \alpha_s (\Sigma^2 -9)}{27 R \sqrt{2}} +  \frac{ 4A \alpha_s^2}{9 R^2 \sqrt{3}}] ( \partial_R^2 + \frac{5}{R} \partial_R ) \phi (R)  \\
  + \frac{8 M^3 \alpha_s \sqrt{2}}{27 R} \phi (R) = 0 $$ 
 
which on rescaling with the dimensionless variable $ X = M R$  leads  to the simpler form 
\begin{equation}\label{5.14}
[1 + \frac{\alpha_s  (\Sigma^2 -9) \sqrt{2}}{24 X} +  \frac{A \alpha_s^2}{\sqrt{3} X^2} ] ( \partial_X^2 + \frac{5}{X}\partial_X 
 ) \phi + \frac{ 2 \alpha_s \sqrt{2}}{3 X} \phi = 0 
\end{equation} 

\subsection{  Spin Effects  of  $V_{qq}$ \& $V_{qqq} $ Terms on $\phi$ Singularity }

The first thing to  notice from the $\phi$ Equation,  (5.14),  is that the spin-dependent parts of both $V_{qq}$ and $V_{qqq}$ appear in the 
multiplying factor with  the 6D 
Laplacian acting on $\phi$, and are proportional to $\alpha_s$ and $\alpha_s^2$ respectively, with   the 
$V_{qq}$  term having  a milder singularity ($ \sim X^{-1}$)  than the  $V_{qqq}$  term ($ \sim X^{-2}$). Further information  on the singularity of 
the differential equation  (at points other than $R=0$)  hinges on the nature of the eigenvalues of the spin operator $A$ which appears in above 
multiplying factor.   Now these eigenvalues  are given by (5.8), two of which are complex for the state of the proton ($ \Sigma^2 = 3$),  but the other two are real. Of the real solutions the positive  eigenvalue does $not$  yield a zero in this multiplying factor  but  the $negative$ eigenvalue $( \approx -4)$ does give a zero for a real  value of $R$  at a point  $ X_0^2$ between $X^2 = 0$ and $X^2 = \infty$ :   
\begin{equation}\label{5.15}
X_0 ^2  - \frac{\alpha_s   \sqrt{2}}{4 }X_0   +  \frac{A \alpha_s^2}{\sqrt{3}} = 0; \nonumber \\
X_0 = + \frac{\alpha_s  \sqrt{2}}{8 } \pm \sqrt{[ \alpha_s^2 / 32  -   \frac{A \alpha_s^2}{\sqrt{3}}]} 
\end{equation}
where the value of $\Sigma^2 =3$ for the proton state has been substituted.  Indeed  this  zero ( for $A \approx -4$)  is a key element of   the  
$dynamical$ effect of the spin - rich  3-body force term, although   the spin effect of the 2-body $V_{qq}$ term is marginal. 
\par
To study  this effect more quantitatively, we seek an approximate solution of Eq. (5.14) in the neighbourhood of the  zero  at  (5.15) 
for which a crude numerical estimate suggests the following location.  Taking a 3-flavour structure $\alpha_s = 2\pi / [9 \ln{M/ \Lambda}]$, 
with $\Lambda \approx 150 MeV $, one finds $\alpha_s \approx 0.39$, whence 
$$ X_0 \approx 0.0689 \pm X_1 ; \quad   X_1 \equiv 0.5841 $$ 
This shows that the $qq$ term gives $\sim 10 \%$ shift around the central value of $X_1$, which corresponds to $ R \approx 0.12 fm$, and 
is almost entirely due to the spin effect of the $V_{qqq}$ term.  Therefore in the spirit of this qualitative investigation,  it makes sense to  
drop this $ 10 \%$ effect, in which case  Eq (5.14) simplifies to                                                           
\begin{equation}\label{5.16}
[1 -  \frac{X_1^2}{ X^2} ] ( \partial_X^2 + \frac{5}{X}\partial_X  ) \phi + \frac{ 2 \alpha_s \sqrt{2}}{3 X} \phi = 0; \quad X_1 = 0.584 
\end{equation} 
To bring it nearer to a standard form, transform to the independent variable $z$ according to  $z X_0^2 = X_0^2 - X^2 $, which yields
\begin{equation}\label{5.17}
z(1-z) \partial_z^2 \phi -3 z \partial_z \phi - \beta \sqrt{1-z} \phi = 0 ; \quad \beta \equiv \frac{\alpha_s}{3\sqrt{2}} X_1 \approx 0.058
\end{equation}   
This equation is almost (not quite) of the hypergeometric form  but it can be reduced to a standard one (with singularities located at  $ z = 0, 1, \infty $) 
[32] by exploiting  the smallness ( $\beta = 0.058$ ) of the last term to replace it with a $constant$   ( with value corresponding to  $ z = 1/2$) :
\begin{equation}\label{5.18}
z(1-z) \partial_z^2 \phi -3 z \partial_z \phi - \frac{\beta}{\sqrt{2}} \phi = 0  
\end{equation}   
An equivalent equation may be obtained with the transformation $ z = 1 - x $ :  
\begin{equation}\label{5.19}
x(1-x) \partial_x^2 \phi + 3 (1-x) \partial_x \phi - \frac{\beta}{\sqrt{2}} \phi = 0  
\end{equation}   
where $x = 1$ corresponds to the location $X = X_1$ of the singularity,  as indicated in Eq. (5.16).   [ Note that this singularity corresponds to  
the $negative$ eigenvalue of $A$, thus reflecting the dynamical effect of the spin-rich $V_{qqq}$ term, while a positive  or complex 
eigenvalue of $A$  would result in the disappearance of this singularity ].   The solution of Eq.(5.19) is then given by [32] 
\begin{equation}\label{5.20}
\phi = F ( a, b | 3 | x) ; \quad a + b = 2; \quad ab = \frac{\beta}{\sqrt{2}} 
\end{equation}
with the entire machinery of hypergeometric functions available [32] for exploring its properties according to  need.  The  scaled variable  
$x$ is related to the 6D distance $R$ by $ MR = X_1 x $ where $ X_1 \approx 1.5 \alpha_s$ is directly related to the location of the singularity 
induced by the spin structure of the Y-shaped  3-body force .  And the reconstruction 
of the full Bethe Salpeter wave function $\Psi$ as a sum of 3 pieces $\Psi_i$ is now only a matter of substituting (5.20) in (4.11), whence the  
corresponding vertex functions $V_i$  are immediately  identified  \cite{Mita99, MitS01} for specific transition amplitudes a la Feynman diagrams. 
  
\section{ Retrospect : Critique Of  The 3-Body Force} 
 
In retrospect,  we have considered, in conjunction with pairwise $qq$ forces,  the effect of a new type  of  3-body force $V_{qqq}$,  rich in spin content,  
 on  the analytical structure of the $qqq$ wave function  in the high momentum regime of QCD  where the confining interaction 
is unimportant, rendering  the dominant force  $Coulombic$. As to the anatomy  of   this (spin-rich)  $V_{qqq}$ ,  we have taken  it to be  generated by   
a $ggg$ vertex ( a genuine part of the  QCD Lagrangian )  wherein  the 3 radiating gluon lines end on  as many quark lines, giving rise to a   
(Mercedes-Benz type)  $Y$-shaped diagram,  a la fig 1. From a physical point of view  it is natural to expect that  such a spin-rich structure    
should play  a   potentially crucial role in  the  so-called `spin anomaly'  of the proton, 
a  subject that  seems once  again  to have raised its head  in the context  of new polarized beam  techniques now available [5] for resolving  
the issue  experimentally.  With  that end in view,  our strategy  has been to determine the $dynamical$  effect of  the spin dependence of   
$V_{qq}$  and $V_{qqq}$ forces 
on the analytical structure of the internal  3D wave function  $\phi$.  [It is emphasized that  this effect  must be carefully  distinguished 
from  the " kinematical "  effect  of spin, which manifests in several other ways, namely through  the presence of various $\gamma$ matrices  
that appear   in   equations  like (4.11) connecting  $\phi$  to the  full 4D  wave function $\Psi$]. Indeed we found in Section 5 that while the 
spin-dynamical effect of  2-body  
forces is marginal, that of the spin-rich 3-body force is quite pronounced,  and shows up   through  the possible " eigenvalues ", Eq (5.7),  of the spin operator  $(i \sigma_1. \sigma_2 \times \sigma_3) $  
which is a part of  $V_{qqq}$ : only a  {\it negative} eigenvalue (there is only one !),   induces  a  {\it singularity}  in the differential equation for $\phi$,  but  not others.  The resulting dynamical effect  is expressed by a hypergeometric function, Eq.(5.20), which gets folded into the 4D  BS amplitude 
$\Psi$ via Eq.(4.11), thus fulfilling the original (limited)  motivation behind this study.   
\par
Unfortunately,  the  dynamical framework underlying the contents of this paper   has   had a long history  born out of  the author's  long  involvement    
with  the so-called Bethe-Salpeter Equation (BSE), often with  extended  periods of stalemate (and frustration !),  but mostly centred around  the  quest for a satisfactory  definition of $probability$ within the BSE framework.  Eventually  it  became possible to  settle for a toned down  version of BSE,  that of a  Salpeter-like equation  ( 3D support for the kernel), which is  amenable to a probability interpretation  at the 3D level.  And  its   
4D features, although present in the original formulation itself \cite{Salp52},  had for decades  remained hidden from view, but  were finally dug out   [7]   
in the context  of an independent approach designed  to explore  a 3D-4D interconnection between the corresponding BS amplitudes when the kernel has a 3D support [8].  Further,   the lack of covariance in the original formulation \cite{Salp52} was 
subsequently remedied via a special (instantaneous) frame of reference in which the composite hadron  of 4-momentum $P_\mu$ is at rest [33]. This 
result  turned out to be  in conformity   with  the Markov-Yukawa Transversality Principle (MYTP) \cite{Mark40, Yuka50}, which  happens       
to be  a `gauge principle'  in disguise \cite{LuOz77} :  It ensures that   the interactions among the constituents  be  $transverse$ to the  
direction of  $P_\mu$.  Subsequent refinements have been  mostly  technical , especially the use of Dirac's  light-front formulation  so as to extend the  
dynamical range of validity of the BS framework  by overcoming  certain practical  problems like `Lorentz mismatch  disease'   \cite{Mitr99} 
associated with different vertices of a given Feynman diagram. 
\par
It therefore looked worthwhile to employ this old-fashioned formalism  in the present context of a new kind of 3-body force $V_{qqq}$,  with enough details 
put in for  a  reasonably  self-contained  description without having to dig frequently into  the original sources. This has also given an opportunity 
to  make several  refinements, especially  the derivation of a common denominator function associated with  both the $qq$ and $qqq$ forces   
and  extending the earlier formalism  to accommodate the $V_{qqq}$ force,  for  which  this  theoretical framework  seems to be well suited.  
\par
The next part of this programme involves  the derivation of the requisite  QCD parameters for  the baryon spin,  for which  some  key ingredients 
are  i) the $forward$ scattering amplitude  off the $i\gamma_\mu \gamma_5$  operator,  inserted at  the  individual quark lines, and ii)   a more 
elaborate  2-gluon  contribution to the proton spin,  which  is reserved for a subsequent paper within  the same  formalism.  

\section{Acknowledgements}

Several colleagues have contributed to the evolution of the  BS   formalism, with and without active authorships, and the present paper also  
shares the general  acknowledgement.  Yet  two items  stand out  in the specific context of this paper :  First,  the  analysis in Section 5 owes its origin  to a  simple-minded  methodology  due to  his  late Father, Jatindranath Mitra.  Second,  the pedagogical techniques of   $qqq$ symmetry with several d.o.f.,s,  which have played a key role in this paper,   are due to  the late Mario Verde  as described in  Hand Buch der Physik Vol 24, 1957.      
The author also acknowledges the comments of Aalok Mishra which led to the use of the eigenvalues of the spin operator $A$ in Section 5. He is also grateful to Vineet Ghildyal for help with figure 1.

\section*{Appendix A : $qq$ Subsystem Formalism}

\setcounter{equation}{0}
\renewcommand{\theequation}{A.\arabic{equation}}

Using the correspondence (2.2) of the text,  most of the covariant instant form  results of \cite{Mita99}   may be taken over to the present light-front situation \cite{MiSo01}. In this Appendix,  we outline the structure of the  4D /  3D interconnection between the corresponding
Green's functions for the $qq$ sub-system as a prototype for the actual $qqq$ system considered in Section 2 of the text. The $qq$ Green's functions satisfy the respective equations \cite{Mita99}:
\begin{equation}\label{A.1}
(2\pi)^4 i G(q,q';P) = \frac{1}{\Delta_1 \Delta_2}\int d^4 q'' 
V({\hat q},{\hat q}'') G(q'',q';P);
\end{equation}
\begin{equation}\label{A.2}
{\hat G}({\hat q},{\hat q}') =  \int {dq_0 dq_0'} G(q,q';P)
\end{equation} 
Integrating both sides of (A.1) gives via (A.2),  the 3D BSE for a 
$bound$ state (no inhomogeneous term !):
\begin{equation}\label{A.3}
(2\pi)^3 D({\hat q}){\hat G}({\hat q},{\hat q}')  
=\int d^3 {\hat q}'' V({\hat q},{\hat q}'') {\hat G}({\hat q}'',{\hat q}')
\end{equation}
where the 3D denominator function $D({\hat q})$ is defined as 
\begin{equation}\label{A.4}
\frac{2i\pi}{D({\hat q})} = \int \frac{dq_0}{\Delta_1 \Delta_2}
\end{equation}
leading (for general unequal mass kinematics) to \cite{CGSM89}
\begin{equation}\label{A.5}
D({\hat q}) = \frac{M}{P_+} D_+({\hat q}); \quad 
D_+({\hat q}) = 2P_+[{\hat q}^2- \frac{\lambda(M^2,m_1^2,m_2^2)}{4M^2}]
\end{equation}
where $\lambda$ is the triangle function of its arguments. Now define the 
hybrid Green's functions \cite{Mita99}:
\begin{equation}\label{A.6}
{\tilde G}({\hat q},q') = \int dq_0 G(q,q';P); \quad 
{\tilde G}(q,{\hat q'}) = \int dq_0' G(q,q';P)
\end{equation}      
Using (A.6) on the RHS of (A.1) gives
\begin{equation}\label{A.7}
(2\pi)^4 i G(q,q';P) = \frac{1}{\Delta_1 \Delta_2}\int d^3 {\hat q}'' 
V({\hat q},{\hat q}'') {\tilde G}({\hat q}'',q')
\end{equation}  
Integrating (A.1) w.r.t. $dq_0'$ only, and using (2.7) again, gives
\begin{equation}\label{A.8}
(2\pi)^4 i {\tilde G}(q,{\hat q}') = \frac{1}{\Delta_1 \Delta_2}\int 
d^3 {\hat q}'' V({\hat q},{\hat q}'') {\hat G}({\hat q}'',{\hat q}')
\end{equation}   
From (A.8) and (A.3), ${\tilde G}$ and ${\hat G}$ are connected as:
\begin{equation}\label{A.9}
{\tilde G}(q,{\hat q}') = \frac{D({\hat q})}{2i\pi \Delta_1 \Delta_2}
{\hat G}({\hat q},{\hat q}')
\end{equation}    
Interchanging $q$ and $q'$ in (A.9) gives the dual result
\begin{equation}\label{A.10}
{\tilde G}({\hat q},q') = \frac{D({\hat q}')}{2i\pi \Delta_1' \Delta_2'}
{\hat G}({\hat q},{\hat q}')
\end{equation}       
Substitution  in (A.7) gives the desired 3D-4D interconnection
\begin{equation}\label{A.11}
G(q,q';P) = \frac{D({\hat q})}{2i\pi \Delta_1 \Delta_2} 
{\hat G}({\hat q},{\hat q}';P) \frac{D({\hat q}')}{2i\pi \Delta_1' \Delta_2'}
\end{equation}
Next, spectral representations for the 4D and 3D G-fns in (A.11) give \cite{Mita99}
\begin{equation}\label{A.12}
G(q,q';P) = \sum_n \Phi_n(q;P) \Phi_n^* (q';P)/ (P^2 +M^2);
\end{equation}
\begin{equation}\label{A.13}
{\hat G}({\hat q},{\hat q}') = \sum_n \phi_n({\hat q})\phi_n^*({\hat q}')
/(P^2+M^2)
\end{equation}
where $\Phi_n$ and $\phi_n$ are 4D and 3D wave functions, so that their interconnection 
(valid for any bound state $n$), is expressed by:
\begin{equation}\label{A.14}
\Gamma({\hat q}) \equiv \Delta_1 \Delta_2 \Phi(q;P)=
\frac{D({\hat q})\phi({\hat q})}{2i\pi}
\end{equation}
which tells us that the covariant vertex function  $\Gamma$ on the light-front is
a function of ${\hat q}$ only \cite{Mita99} via its definition (2.2).  This derivation is a prototype for the $qqq$ case  of text.

\end{document}